\documentclass[
reprint,
altaffilletter,
nofootinbib,
amsmath,
amssymb,
aps,
prd]{revtex4-1}

\usepackage{mathtools}
\usepackage{graphicx}
\usepackage{dcolumn}
\newcolumntype{d}[1]{D{.}{.}{#1}}
\usepackage{bm}
\usepackage{soul}
\usepackage{verbatim} 
\usepackage{cprotect} 
\usepackage{url}
\usepackage{subfigure} 
\usepackage{tikz}
\usepackage{afterpage}
\usepackage{nicefrac}
\usepackage{tabularx}
\newcolumntype{L}{>{\raggedright\arraybackslash}X}
\usepackage{multirow}
\usepackage[english]{babel}
\usepackage{natbib} 
\usepackage{glossaries} 
\usepackage{aas_macros} 
\usepackage{comment}
\usepackage{float}
\usepackage{placeins}
\usepackage{ulem}[normalem]
\usepackage{array}

\graphicspath{{figures/}}
\usepackage{hyperref}
\hypersetup{
	colorlinks=true,
	citecolor=NavyBlue,
	linkcolor=NavyBlue,
	urlcolor=NavyBlue,
}

\usepackage[dvipsnames]{xcolor} 

\newcolumntype{C}[1]{>{\arraybackslash}p{#1}}

\newcommand{\ba}{\begin{eqnarray}}
\newcommand{\ea}{\end{eqnarray}}
\newcommand{\be}{\begin{equation}}
\newcommand{\ee}{\end{equation}}

\newcommand{\murel}{\mu_{\text{rel}}}

\definecolor{grey}{rgb}{0.4,0.4,0.4}
\definecolor{dullmagenta}{rgb}{0.4,0,0.4}
\definecolor{darkblue}{rgb}{0,0,0.4}
\definecolor{midblue}{rgb}{0,0,0.5}
\definecolor{midred}{rgb}{0.5,0,0}
\definecolor{orange}{rgb}{1,0.5,0}
\definecolor{lightbrown}{rgb}{0.75,0.5,0.25}
\definecolor{tan}{cmyk}{0.14,0.42,0.56,0}
\definecolor{djunglegreen}{cmyk}{0.99,0,0.52,0}
\definecolor{lightgreen}{rgb}{0,1,0}
\definecolor{olivegreen}{cmyk}{0.64,0,0.95,0.40}
\definecolor{midgreen}{rgb}{0.0,0.675,0.0}
\definecolor{darkgreen}{rgb}{0,0.5,0}

\newcommand{\dingolensing}{\texttt{DINGO-lensing}}
\newcommand{\dingo}{\texttt{DINGO}}
\newcommand{\modwaveforms}{\texttt{modwaveforms}}
\newcommand{\bilby}{\texttt{bilby}}
\newcommand{\Dt}{\Delta t}

\newcommand{\BFlens}{\mathcal{B}_\mathrm{lens}}
\newcommand{\Dlens}{\mathcal{D}_\mathrm{nonlens}}
\newcommand{\ForegroundLargerFive}{53\%}
\newcommand{\BackgroundLargerZero}{2\%}
\newcommand{\Nforeground}{2,561}
\newcommand{\Nbackground}{10,858}
\usepackage{orcidlink}

\begin{document}

\title{Identification and characterization of distorted gravitational waves by lensing \\ using deep learning}

\author{Juno C. L. Chan %\orcidlink{0000-0002-3377-4737}
}
\email{chun.lung.chan@nbi.ku.dk}
\author{Lorena \surname{Maga\~na Zertuche} %\orcidlink{0000-0003-1888-9904}
}
\email{lorena.zertuche@nbi.ku.dk}
\author{Jose Mar\'ia Ezquiaga %\orcidlink{}
}
\author{Rico K.~L.~Lo %\orcidlink{}
}
\author{Luka Vujeva %\orcidlink{0000-0001-7697-8361}
}
\author{Joey Bowman %\orcidlink{https://orcid.org/0009-0003-3582-1984}
}

\affiliation{Center of Gravity, Niels Bohr Institute, Blegdamsvej 17, 2100 Copenhagen, Denmark}

\date{\today}

\begin{abstract}
Gravitational waves (GWs) can be distorted by intervening mass distributions while propagating, leading to frequency-dependent modulations that imprint a distinct signature on the observed waveforms. Bayesian inference for GW lensing with conventional sampling methods is costly, and the problem is exacerbated by the rapidly growing GW catalog. Moreover, assessing the statistical significance of lensed candidates requires thousands, if not millions, of simulations to estimate the background from noise fluctuations and waveform systematics, which is infeasible with standard samplers. We present a novel method, \dingolensing, for performing inference on lensed GWs, extending the neural posterior estimation framework \dingo. By comparing our results with those using conventional samplers, we show that the compute time of parameter estimation of lensed GWs can be reduced from weeks to seconds, while preserving accuracy both in the posterior distributions and the evidence ratios. 
We train our neural networks with LIGO detector noise at design sensitivity and a lens model that accommodates two overlapping images with a constant $\pi/2$ phase shift. 
We show that the lensing parameters are recovered with millisecond precision for the time delays. We also demonstrate that our network can identify signals diffracted by point masses, highlighting its flexibility for searches. By simulating thousands of lensed and nonlensed events, we determine how the detectability changes with different source properties. \dingolensing{} provides a scalable and efficient avenue for identifying and characterizing gravitationally lensed GW events in the upcoming observing runs.
\end{abstract}

\maketitle

%==========================================================
\section{Introduction}\label{Section: Introduction}
%==========================================================

The strongest emission of gravitational waves (GWs) in the Universe comes from compact binary coalescences (CBCs), such as binary neutron stars, neutron star-black hole systems, and binary black holes (BBHs). The signals coming from these highly energetic events are detected by the LIGO-Virgo-KAGRA (LVK) detector network~\cite{LIGOScientific:2014pky,VIRGO:2014yos,KAGRA:2020tym}, which has already brought us many confident BBH GW detections~\cite{GWTC4}. By comparing the observed signals to those coming from numerical relativity or a robust model, we are able to estimate the intrinsic and extrinsic parameters of such binary systems. In order to do this accurately, we rely on cutting-edge models that capture several physical effects, such as higher harmonic modes, eccentricity, and precession~\cite{Field:2013cfa,Nagar:2018zoe,Varma:2018mmi,Varma:2019csw,Pratten:2020ceb,Ossokine:2020kjp,Hamilton:2021pkf,Pompili:2023tna,Ramos-Buades:2023ehm,Thompson:2023ase,Colleoni:2024knd,Albanesi:2025txj}. 

GWs propagate over cosmological distances and are not impervious to intervening matter. Instead, their path is affected by nearby gravitational potentials (acting as lenses), such as those of stars, black holes, galaxies, and clusters of galaxies, leading to gravitational lensing~\cite{Nakamura:1997sw, Takahashi:2003ix}. In strong gravitational lensing, multiple signals, or images, of the source are produced each with a distinct amplitude, phase shift, and time delay~\cite{Wang:1996as, Dai:2017huk, Ezquiaga:2020gdt}. The arrival time difference between these images can be very small for smaller lens masses (or specific source-lens-observer configurations)~\cite{Schneider:1992bmb,Lo:2024wqm,Vujeva:2025kko,Ezquiaga:2025gkd,Vujeva:2025nwg}, making it challenging to look for signals coming from the same source. Additionally, these small lens masses may produce overlapping images which causes a beating pattern in the inspiral portion of the waveform resembling precession or the effect of two different binary signals overlapping~\cite{Liu:2023emk,Rao:2025poe}. 
When the GW wavelength is comparable to or larger than the characteristic scale of the lens, diffraction becomes significant and a single signal can exhibit frequency-dependent modulations. 
Detecting such signal can be challenging in matched-filtering searches~\cite{Chan:2024qmb},
but it can answer key questions in the small lens mass regime regarding small compact lens populations~\cite{Jung:2017flg,Lai:2018rto,Christian:2018vsi,Dai:2018enj,Diego:2019lcd,Diego:2019rzc,Liao:2020hnx,Mishra:2021xzz,Seo:2021psp,Seo:2023rjd,Liu:2023ikc,Cheung:2024ugg,Zumalacarregui:2024ocb,Chakraborty:2024mbr,Chan:2025wgz}, the halo mass function~\cite{Fairbairn:2022xln,Tambalo:2022wlm,Caliskan:2022hbu,Savastano:2023spl}, the structures of different lenses~\cite{Lo:2024wqm,Vujeva:2025kko,Ezquiaga:2025gkd,Vujeva:2025nwg,Kim:2025njb,Abe:2025ahi,Ali:2025cqo,Su:2025xry,Bulashenko:2025vdx,Singh:2025uvp}, the cosmic expansion rate~\cite{Cremonese:2021puh,Chen:2024xal,Smith:2025axx}, and the astrophysical environments of GW sources~\cite{Leong:2024nnx,Li:2025xuh,Ubach:2025dob}, as well as provide an alternate way to test for deviations of general relativity~\cite{Goyal:2020bkm,Goyal:2023uvm,Narola:2023viz,Ezquiaga:2020dao,Liu:2024xxn}.

Confirming the lensed nature of a candidate event requires scrutiny, even in small effects such as noise fluctuations and waveform systematics ~\cite{Keitel:2024brp}.
Making a confident detection claim requires the simulation and analysis of millions of events. 
Carefully conducted searches for lensing signatures in GWs in the first three  
observing runs found no convincing evidence for GW lensing~\cite{Hannuksela:2019kle,Dai:2020tpj,LIGOScientific:2021izm,LIGOScientific:2023bwz,Janquart:2023mvf,Chakraborty:2025maj}. However, the fourth observing run brought a lensed candidate, GW231123~\cite{GW231123}, which we investigate in detail in a companion paper~\cite{Chan:2025kyu}.  

The high computational cost of performing Bayesian inference on distorted GW signals with conventional sampling algorithms, such as Markov chain Monte Carlo and nested sampling~\cite{Thrane:2018qnx,Liang:2025yjw}, poses a major limitation for lensing studies. 
Not only does it prevent us from keeping pace with the rapidly growing number of detections expected from an expanding detector horizon, but it also makes background estimation for lensing searches prohibitively expensive, as millions of nonlensed parameter estimation (PE) runs are required to quantify the lensing false-alarm rate. 
To address these challenges, several alternatives that bypass conventional Bayesian inference have been proposed~\cite{Seo:2025dto,Qin:2025mvj}. 
While these methods offer significant speedups, they often lack the full statistical interpretability and robustness of standard Bayesian inference.

Deep-learning techniques have emerged as a powerful solution for performing fast and accurate PE of GW signals~\cite{Liang:2025yjw}. 
Unlike common Bayesian sampling methods---which require repeated waveform evaluations and are therefore computationally expensive---neural networks can learn the complex mapping between GW strain data and source parameters directly from simulated training sets. 
Once trained, these models can generate full posteriors within seconds, achieving orders-of-magnitude speedups while maintaining precision comparable to traditional PE pipelines~\cite{Chua:2019wwt,Gabbard:2019rde,Cuoco:2020ogp,Langendorff:2022fzq,Zhao:2023tqr,Bhardwaj:2023xph,Cuoco:2024cdk,Hu:2024oen,Polanska:2024zpn,Stergioulas:2024jgk,Papalini:2025exy,Santoliquido:2025lot}.

%Recent studies have begun extending deep-learning techniques to the lensed GW regime, exploring the use of convolutional-, recurrent-, and transformer-based architectures for rapid identification of lensing signatures~\cite{Kim:2020xkm,Goyal:2021hxv,Magare:2024wje} and inferring lens parameters~\cite{Bada-Nerin:2024wkn,Qin:2025mvj}. 
Recent studies have begun extending deep-learning techniques for the rapid identification of strongly lensed GW signals~\cite{Goyal:2021hxv,Magare:2024wje}.
In the wave-optics regime, several architectures have been explored, including visual geometry group networks~\cite{Kim:2020xkm}, conditional variational autoencoders~\cite{Bada-Nerin:2024wkn}, and neural spline flows~\cite{Qin:2025mvj}.
Reference~\cite{Kim:2020xkm} presents a proof-of-principle demonstration that deep-learning models can identify waveform distortions.
Reference~\cite{Bada-Nerin:2024wkn} further shows a substantial speed-up in inferring lens parameters compared to traditional Bayesian sampling.
Reference~\cite{Qin:2025mvj} extends the inference dimensionality to eleven.
Among these approaches, however, a rapid inference of all source parameters with accuracy comparable to traditional Bayesian sampling methods has not yet been achieved.

The \dingo{} framework~\cite{Green:2020hst,Green:2020dnx,Dax:2021tsq,Dax:2021myb,Wildberger:2022agw,Dax:2022pxd,Gupte:2024jfe,Santoliquido:2025lot,Hu:2025vlp} represents a state-of-the-art implementation of simulation-based inference using neural posterior estimation (NPE). 
\dingo{} combines an embedding network with normalizing flows to produce full Bayesian posteriors conditioned on GW data, enabling rapid and high-fidelity inference across high-dimensional parameter spaces~\cite{Green:2020dnx,Dax:2021tsq}. Simulation-based inference provides a solution by combining computational efficiency and statistical interpretability.
Instead of relying on explicit likelihood evaluations, simulation-based inference employs neural density estimators trained on simulated data to learn the complex mapping between observations and parameters~\cite{Cranmer:2019eaq,Liang:2025yjw}. 
This offers a promising deep-learning pathway toward rapid and accurate inference on lensed GWs.

Building on top of the simulation-based inference algorithm, \dingo, we present a new lensing inference pipeline \dingolensing\footnote{\href{https://github.com/dingo-lensing/dingo-lensing}{https://github.com/dingo-lensing/dingo-lensing}}.
In particular, we employ a flexible two image model in our lensed waveform generation that is able to accommodate the most common wave-optics models~\cite{Ezquiaga:2025gkd}.
By directly learning the posterior distribution from simulated data, \dingolensing{} circumvents the computational bottlenecks of a traditional sampler while maintaining the statistical rigour of conventional inference. 
This approach establishes a scalable and robust pathway for identifying and characterizing lensed GW events in the era of high-cadence detections.

This paper is structured as follows: 
Section~\ref{Section: Methods} describes the lensed waveform model, network architecture, and training setup.
Section~\ref{Section: Results} presents the results and validation of our trained network, including comparisons with traditional Bayesian inference, systematics, and an assessment of detectability.
Finally, Section~\ref{Section: Conclusion} summarizes our findings and discusses the implications for future applications of simulation-based inference in lensing analyses.

%==========================================================
\section{\dingolensing }\label{Section: Methods}
%==========================================================

In this work, we perform NPE training with the simulation-based inference framework \dingo. 
We use the fifth observing run (O5) noise amplitude spectral density (ASD)~\cite{KAGRA:2013rdx,sensitivity_curves_ligo}, 
assuming Gaussian and stationary noise.
Since glitches cause the noise to deviate from the Gaussian and stationary assumptions used here, we will assess their influence in future studies.
Throughout the study, we utilize the inspiral-merger-ringdown phenomenological model with higher modes and precession, \texttt{IMRPhenomXPHM}~\cite{Pratten:2020ceb} implemented in \texttt{LALSimulation}~\cite{lalsuite, swiglal}, with a frequency range of $[20, 1024]~\rm{Hz}$ and a waveform duration of $8~\rm{s}$. We have modified the network to allow for the use of numerical relativity surrogate waveforms during training and parameter inference, e.g., \texttt{NRSur7dq4} that spans a frequency range of [20, 512] Hz, which we use in the companion paper~\cite{Chan:2025kyu}.
The intrinsic parameters consist of 
the component masses in the detector frame $m_i~(i=1,2)$,
the spin magnitudes $a_i$, and 
the spin angles $(\phi_i,\phi_{12},\phi_{JL})$.
Other parameters included are 
the inclination angle $\theta_{JN}$, 
right ascension (RA) and declination (DEC), 
polarization angle $\psi$,
luminosity distance $d_{\rm L}$, the time of the peak amplitude at geocenter $t_{\rm c}$, and the phase at a reference frequency $\phi_{\rm ref}$ of 20Hz.

The following subsections present the lensed waveform model adopted in constructing the training dataset and in the subsequent analyses, along with a description of the training setup.

%==========================================================
\subsection{Lensed waveform model}
%==========================================================

With the weak-gravity and thin-lens approximation, the gravitational lensing of a GW signal $h(t)$ can be solved efficiently in the frequency domain~\cite{Nakamura:1997sw,Takahashi:2003ix}. 
The lensed $\tilde h_{\rm L}(f)$ and unlensed $\tilde h_{\rm U}(f)$ frequency-dependent GW signals are related by the amplification factor $F(f)$ as
\begin{equation}
    \tilde h_{\rm L}(f) = F(f)\tilde h_{\rm U}(f)\,.
\end{equation}
The amplification factor is obtained by solving the diffraction integral~\cite{Takahashi:2003ix},
\begin{equation}
    F(f,\boldsymbol{\theta}_{\rm S})=\frac{\tau_{D}f}{i}\int \mathrm{d}^2\boldsymbol{\theta}_{\rm{I}}\, e^{2\pi i f  t_{\rm d}(\boldsymbol{\theta}_{\rm{I}},\boldsymbol{\theta}_{\rm S})}\,,
\end{equation}
which integrates over the image plane positions $\boldsymbol{\theta}_{\rm{I}}$. 
Here, 
$\tau_{D}$ is the characteristic timescale associated to the distances between source-lens-observer, $\boldsymbol{\theta}_{\rm S}$ is the nonlensed source angular position, and $t_{\rm d}(\boldsymbol{\theta}_{\rm{I}},\boldsymbol{\theta}_{\rm S})$ is the time delay between the source and observer, often called the \textit{time-delay surface}.
The notations follow the definition in~\cite{Ezquiaga:2025gkd}.

Using the stationary phase approximation (SPA), the diffraction integral simplifies to a sum over multiple copies of the original signal. 
Each of these copies can be magnified, time delayed and phase shifted. 
If the time delay between any of the repeated copies is less than the signal's duration, then there are interference effects that distort the signal. 
In this work, we focus on  
an amplification factor that depends on the time delay $\Dt$ and relative magnification $\murel$ between two images with a phase difference of $\pi/2$, i.e.,
\begin{equation}
\label{eq:F_two_images_point}
F(f>0) = 1 + \sqrt{\murel }\, e^{i(2\pi f \Dt - \pi/2)}\,.
\end{equation}
This expression, which is equivalent to the strong-lensing regime~\cite{Dai:2017huk,Ezquiaga:2020gdt}, is a good approximation for systems in which two images are interfering with each other and  
has been shown to hold for several types of singularities, including point, fold, and (some) cusp cases (see ~\cite{Serra:2025kbw,Ezquiaga:2025gkd} where further details on its validity can be found). 
Operationally, we implement this lens model coded in \modwaveforms.\footnote{\href{https://github.com/ezquiaga/modwaveforms}{https://github.com/ezquiaga/modwaveforms}}

Image properties are determined by their locations on the time delay surface. 
That is, images that form on minima of the time delay surface are of type I with no phase shift, those that form on saddle points are of type II with a $\pi/2$ phase shift, and those on maxima are of type III with a $\pi$ phase shift~\cite{Blandford:1986zz}. 
For example, a point-mass lens always produces two images of type I and II~\cite{Schneider:1992bmb}.
For the point, fold, and many configurations near cusp singularities, the brightest image (of type I) arrives before the second brightest image (of type II)~\cite{Blandford:1986zz}. 
Because of this hierarchy in the magnifications, this allows us to define our relative magnification, or the ratio of the second brightest to the brightest image $\murel \equiv \mu_2/\mu_1$, as being restricted to $\murel\in[0,1]$ and the time delay between the second and first image as $\Dt> 0$. Because the type I image arrives before the type II image, this will always result in a $\pi/2$ phase difference between the two lensed images. 
This assumption removes a possible degeneracy where images of the same type interfering with each other can lead to $(\murel, \Dt) = (0,0)$ or $(\murel, \Dt) = (1,0)$, which is broken in our case due to the phase difference between the two images. Therefore, within our framework we define a ``lensed image'' as an event in which our analysis finds $\Dt>0$, implying $\murel<1$. 
We define a ``nonlensed event'' as an event which our analysis finds a relative magnification factor of $\murel = 0$.  
Exploring lensing configurations where one would find overlapping images of the same type, as well as images with $\murel>1$ such as the ones created by lenses containing substructure \cite{Vujeva:2025kko, Vujeva:2025nwg}, is left to future work. 

\begin{figure}
    \centering
    \includegraphics[width=\columnwidth]{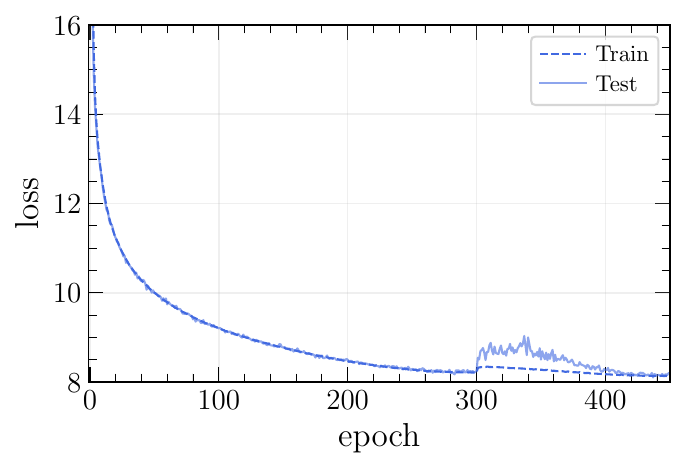}
    \caption{The loss as a function of the number of training epochs of the lensed network, shown for both the training set (dashed) and testing set (solid).}
    \label{logloss}
\end{figure}

%==========================================================
\subsection{Training setup}
%==========================================================

We perform GW parameter estimation using \dingolensing, building on top of the inference pipeline \dingo~\cite{Green:2020hst,Green:2020dnx,Dax:2021tsq,Dax:2021myb,Wildberger:2022agw,Dax:2022pxd}.
We use NPE to model the Bayesian conditional distribution probability $q_{\phi}(\boldsymbol{\theta} | d)$ as a surrogate for the true posterior probability $p(\boldsymbol{\theta} | d)$, given by
\begin{equation}
\label{eq:bayes_theorem}
p(\boldsymbol{\theta} | d) = \frac{\mathcal{L}(d | \boldsymbol{\theta}) \pi(\boldsymbol{\theta})}{\mathcal{Z}(d)},
\end{equation}
where $\phi$ are the training parameters in the network. 
Training is carried out on simulated, labeled pairs ($\boldsymbol{\theta}, d$), where the parameters $\boldsymbol{\theta}$ are drawn from the prior $\pi(\boldsymbol{\theta})$, and the corresponding detector strain $d$ is sampled from the likelihood $\mathcal{L}(d | \boldsymbol{\theta})$. 
In practice, this involves injecting a waveform $h(\boldsymbol{\theta}_i)$ into additive Gaussian noise $n_i$, assuming the noise is stationary: $d_i = h(\boldsymbol{\theta}_i) + n_i$. 

The network architecture follows the design in~\cite{Dax:2021myb}, with all hyperparameters kept unchanged. 
In short, \dingolensing{} employs a dual-network structure: an embedding network, whose first layer reduces the raw strain data to a 128-dimensional representation based on singular value decomposition (SVD), followed by a fully connected network that maps this representation into an approximation of the posterior. Our architecture comprises $\sim 10^8$ learnable parameters and $\sim 10^7$ fixed parameters.

\begin{table}[htbp]
\centering
\begin{tabular}{ccc|c}
\hline
\textbf{Parameter} & \textbf{Unit} & \textbf{Prior} & \textbf{Injection} \\ 
\hline
$\mathcal{M}_{\rm c}$ & $M_\odot$ & $\mathcal{U}(10, 100)$ & 49.2 \\
$q$ & $M_\odot$ & $\mathcal{U}(0.125, 1.0)$ & 0.79 \\
$m_1$  & $M_\odot$& $\mathcal{C}(5, 150)$ & 63.6\\ 
$m_2$  & $M_\odot$& $\mathcal{C}(5, 150)$ & 50.4\\
$a_1$ & & $\mathcal{U}(0, 0.99)$ & 0.28 \\
$a_2$ & & $\mathcal{U}(0, 0.99)$ & 0.45 \\
 $\phi_{12}$ & & $\mathcal{U}(0, 2\pi)$ & 3.59 \\
$\phi_1$ &  & $\sin{(0, \pi)}$ & 1.75\\
$\phi_2$ &  & $\sin{(0, \pi)}$ & 0.49\\
$\phi_{JL}$ &  & $\mathcal{U}(0, 2\pi)$ & 5.38\\
$\theta_{JN}$ &  & $\sin{(0, \pi)}$ & 2.83\\
$\phi_{\rm ref}$ & & $\mathcal{U}(0, 2\pi)$ & 1.73\\
RA & & $\mathcal{U}(0,2\pi)$ & 3.01\\
DEC & & $\cos(-\pi/2, \pi/2)$  & -0.13 \\
$\psi$ & & $\mathcal{U}(0,\pi)$ & 2.63\\
$t_{\rm c}$ & $\mathrm{s}$ & $\mathcal{U}(-0.1, 0.1)$ & $-0.0565$ \\
$d_{\rm L}$ & $\mathrm{Gpc}$ &  $\mathcal{U}(1,15)$ & 9.84\\
\hline
$\Delta t$ & $\mathrm{s}$ & $\mathcal{U}(0, 0.1)$ & 0.031\\
$\murel$ & & $\mathcal{U}(0, 1)$ & 0.79\\
\hline
\end{tabular}
\caption{The first three columns show the prior distributions corresponding to a given parameter used in the training dataset for NPE, where $\mathcal{U}$ denotes a uniform distribution and $\mathcal{C}$ denotes a parameter is constrained to be in such range. 
The last column has the injected values for comparing the accuracy of \dingolensing{} and \bilby{} in Figs.~\ref{corner} and \ref{corner_nonlensed}.}
\label{tab:priors}
\end{table}

We generate training datasets consisting of $5 \times 10^{6}$ simulated waveforms, with parameters drawn from the priors listed in Table~\ref{tab:priors}. The waveforms are stored in an SVD representation, and $2\%$ of the dataset is reserved for testing.

\begin{table*}[t]
\centering
\begin{tabular}{l @{\hspace{0.5cm}} c @{\hspace{0.5cm}} c}
\hline\hline
\textbf{Training setting} & \textbf{Stage 0} & \textbf{Stage 1} \\
\hline
Layers & SVD layer frozen & All layers unfrozen \\
Epochs & 300 & 150 \\
Optimizer & Adam~\cite{kingma2014} & Adam \\
Learning rate & $7\times10^{-5}$ & $1\times10^{-5}$ \\
Scheduler & Cosine scheduler~\cite{2016arXiv160803983L} & Cosine scheduler \\
Batch size & 4096 & 4096 \\
Temperature & 300 & 150 \\
Purpose & Stabilize early training & Fine-tuning for accuracy \\
\hline\hline
\end{tabular}
\caption{Summary of the two-stage NPE training scheme.}
\label{table_training}
\end{table*}

We adopt the two-stage training scheme with the hyperparameters summarized in Table~\ref{table_training}.
The initial training stage (Stage $0$) freezes the initial SVD layer and is unfreezed in the fine-tuning stage (Stage $1$).
The training of both the lensed and nonlensed NPE models is performed in two stages. 
During Stage 0, the model is trained for 300 epochs using the Adam optimizer \cite{kingma2014} with a learning rate of $7\times10^{-5}$ and a cosine learning-rate scheduler \cite{2016arXiv160803983L}. 
In Stage 1, all layers are unfrozen for fine-tuning over an additional 150 epochs, with the learning rate reduced to $1\times10^{-5}$. 
Both stages employ a batch size of 4096 and temperatures of 300 and 150, respectively. 
This two-stage procedure stabilizes early optimization and enables the network to refine the embedding representation for improved posterior accuracy.

During the training process, a neural network evaluates its performance by computing a \textit{loss function}, which measures the discrepancy between the predicted and target probability distributions. 
The loss function corresponds to the negative log-likelihood of the data, under the forward Kullback–Leibler divergence model, between the true posterior $p(\boldsymbol{\theta}|d)$ and the neural approximation $q_\phi(\boldsymbol{\theta}|d)$~\cite{Green:2020dnx}.

Minimizing this loss encourages the network to produce posteriors that closely match those obtained from exact Bayesian inference. The optimization proceeds iteratively over epochs by adjusting the network parameters $\phi$, typically via stochastic gradient descent, until convergence, as exemplified in Figure~\ref{logloss}.

Note that we tested training configurations with different batch sizes, temperatures, and learning rates.
While our network validation in Section~\ref{Section: Results} focuses on the setup presented in Table~\ref{table_training}, we find that the choice of batch size and temperature introduces negligible variation in performance.
A smaller learning rate, however, provides more stable training and better performance when the range of luminosity distance is increased.

\subsection{Importance sampling}
To obtain accurate posterior estimates without retraining the neural network for each new likelihood configuration, \dingo{} employs importance sampling as a correction step following the NPE stage \cite{Dax:2022pxd}. Given a trained model that approximates the posterior distribution $q_\phi(\boldsymbol{\theta}|d) $ over source and lensing parameters $\boldsymbol{\theta}$ conditioned on data $d$, we reweight samples from this approximate posterior to the true target posterior under a desired likelihood $p(d|\boldsymbol{\theta})$ and prior $p(\boldsymbol{\theta})$.

The corrected posterior is obtained via the standard importance reweighting relation:
\begin{equation}
    p(\boldsymbol{\theta}|d) \propto 
\frac{p(d|\boldsymbol{\theta})p(\boldsymbol{\theta})}
     {q_\phi(\boldsymbol{\theta}|d)}.
\end{equation}
Given a set of $N$ samples $\{\boldsymbol{\theta}_i\}_{i=1}^N$ drawn from $q_\phi(\boldsymbol{\theta}|d)$, the corresponding importance weights are computed as
\begin{equation}
    w_i = \frac{p(d|\boldsymbol{\theta}_i)\,p(\boldsymbol{\theta}_i)}
            {q_\phi(\boldsymbol{\theta}_i|d)},
\qquad 
\tilde{w}_i = \frac{w_i}{\sum_{j=1}^{N} w_j},
\end{equation}
where \(\tilde{w}_i\) are the normalized weights satisfying $\sum_i \tilde{w}_i = 1$.  
Expectations of any quantity $f(\boldsymbol{\theta})$ under the target posterior are then estimated as
\begin{equation}
    \mathbb{E}_{p}[f(\boldsymbol{\theta})]
  \approx \sum_{i=1}^{N} \tilde{w}_i\, f(\boldsymbol{\theta}_i).
\end{equation}

In practice, the importance weights are computed using the likelihood function implemented in \bilby{}, ensuring that the reweighted samples are consistent with the full noise model and data segment used for parameter estimation.  
This step allows the neural posterior to flexibly adapt to small mismatches between the simulation-based training distribution and the actual data likelihood, without retraining the network.  
The efficiency of this procedure is monitored via the sampling efficiency,
\begin{equation}
    \epsilon = \frac{1}{N} \frac{\left(\sum_i \tilde{w}_i\right)^2}{\sum_i \tilde{w}_i^2},
\end{equation}
which quantifies the effective number of posterior samples contributing to the reweighted estimate.  
In our analyses, $\epsilon$ typically remains above a few percent, indicating that the learned posterior provides a sufficiently close approximation to the true one for reliable reweighting.

\subsection{Lensing Bayes factor}\label{lens_BF}
To quantify the support for the lensed hypothesis over the not lensed hypothesis, we compute the Bayes factor
\begin{equation}
    \BFlens = \frac{p(d | \mathcal{H}_\mathrm{L})}{p(d | \mathcal{H}_\mathrm{NL})},
\end{equation}
where $\mathcal{H}_\mathrm{L}$ and $\mathcal{H}_\mathrm{NL}$ denote the lensed and not lensed hypotheses, respectively.

The Bayesian evidence under each hypothesis is approximated via importance sampling. Specifically, samples are drawn from the neural posterior estimator and reweighted according to the true likelihood-to-network ratio, following the approach of~\cite{Dax:2022pxd}.
This procedure allows for an efficient and unbiased estimate of the marginal likelihood without requiring explicit integration over the full parameter space. 
We train separate nonlensed and lensed models to perform posterior estimation under each hypothesis with the training setup and validation described in Appendix~\ref{Appdx:not lensed}.

%==========================================================
\section{Results}\label{Section: Results}
%==========================================================

In this section, we present the training results of our lensed network and summarize its performance through a series of model validation tests.
We further compare representative corner plots with \bilby\ and demonstrate that \dingolensing~can effectively recover point-lens injections.

\subsection{Training performance}
Figure~\ref{logloss} shows the evolution of the loss as a function of training epoch. 
The training loss (dashed) and test loss (solid) remain closely aligned throughout, indicating the absence of overfitting. 
The smooth behavior of both curves further demonstrates the stability of the training process. 
After approximately $400$ epochs, the loss function reaches a plateau, suggesting that the model has ceased to learn significant new information from the data. 
A noticeable discontinuity around epoch $300$ corresponds to unfreezing the additional layers in stage~$1$. 
We terminate the training at epoch~$450$, as no further improvement in accuracy is observed.
The training takes $\sim 10~\rm{days}$ with an NVIDIA A100 PCIe GPU. 

\begin{figure}
    %\centering
    \includegraphics[width=0.9\columnwidth]{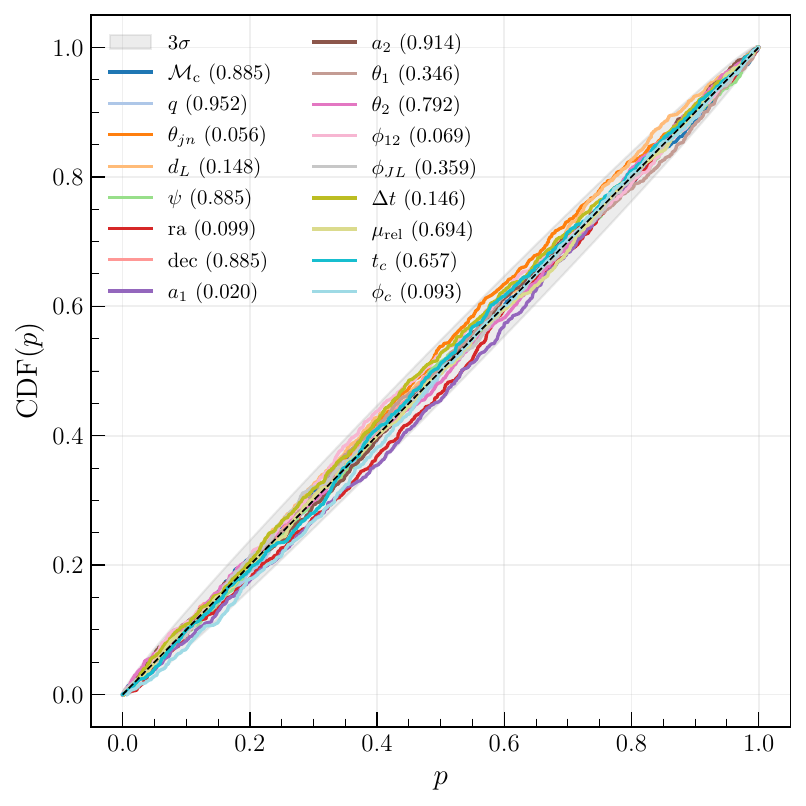}
    \caption{P-P plot for the lensed neural posterior estimation model using $10^3$ lensed injections and no importance sampling.
    For each injection, we generate a posterior with our model and compute the percentile value of the injected parameter.
    Each colored line corresponds to the cumulative distribution function (CDF)
    of the corresponding parameters (see label).
    The Kolmogorov-Smirnov test $p$-values are given in the legend.
    }
    \label{pp_plot_lensed}
\end{figure}

\begin{figure*}[t!]
    \centering
    \includegraphics[width=\textwidth]{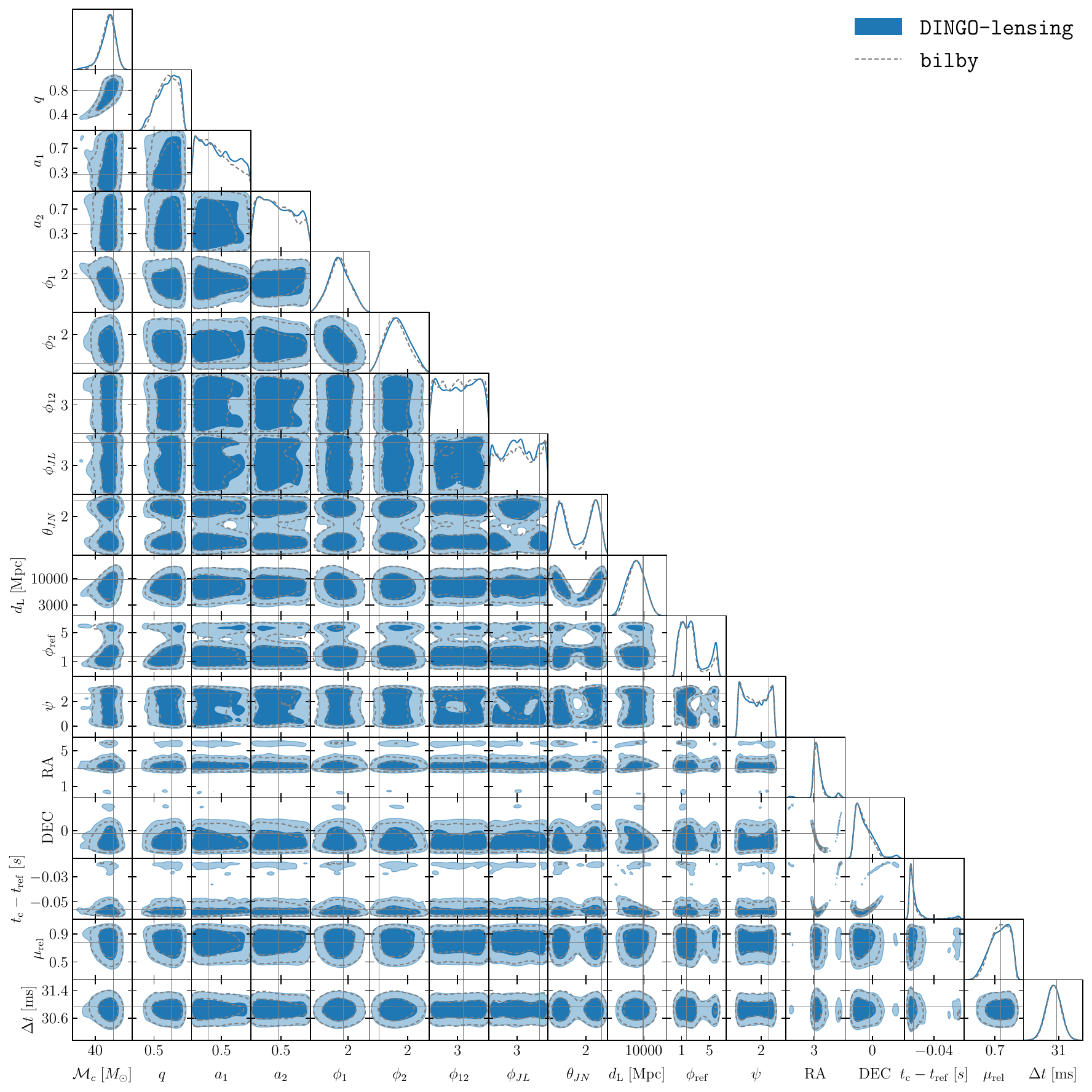}
    \caption{Posterior distribution of an injection recovered using our \dingolensing{} (solid blue curves) and \bilby{} (dashed gray curves), respectively. The two methods agree with each other and both recovered the injected values (solid gray lines; also tabulated in Tab.~\ref{tab:priors}) well.}
    \label{corner}
\end{figure*}

\subsection{Network validation}

We perform inference on $10^3$ lensed injections generated from our trained model and draw $10^4$ samples for each injections from the lensed network. 
We then reconstruct the reference phase \cite{Dax:2022pxd}. 
Figure~\ref{pp_plot_lensed} displays the corresponding probability–probability (P–P) plot constructed directly from the posteriors obtained by the lensed network without applying importance sampling. 
The diagonal line represents the ideal case in which $x\%$ of the true parameters lie within the $x\%$ credible intervals of the inferred posteriors. 
The colored curves for individual parameters show excellent agreement with this diagonal, demonstrating that the network produces well-calibrated posteriors. 
This proper calibration confirms that the sampling procedure functions as intended, a result further supported by the Kolmogorov–Smirnov test $p$-values shown in the legend~\cite{seabold2010statsmodels}. 
Minor deviations for the lensing parameters ($\Dt$ and $\murel$) remain within expected statistical fluctuations ($3\sigma$), indicating robust performance of the model even in the presence of lensing effects.

In addition, we look at the distribution of sampling efficiencies $\epsilon$ after importance sampling. 
We find that they span a wide range of values between $\sim10^{-4}$ and $\sim0.1$ with no particular correlation with the binary and lens parameters. 
Moreover, we have tested the effects of sampling efficiency on the Bayes factor computation and find good agreement between \dingolensing{} and \bilby{} for $\epsilon>10^{-4}$. 
For more details see Appendix \ref{app:BF_sampling_efficiency}.

\subsection{Comparison with traditional parameter estimation methods}
\label{subsec:comparison_with_bilby}
Figure~\ref{corner} shows a corner plot comparing the posterior distributions obtained using our \dingolensing{} model (solid blue curves) with the reference \bilby{} inference pipeline (dashed gray curves) for a representative lensed injection with a network SNR~$=16.18$ at O5 sensitivity. 
The source parameters and priors used in the \bilby{} analysis are tabulated in Table~\ref{tab:priors}.

The one- and two-dimensional marginalized posteriors show excellent agreement between \dingolensing{} and \bilby{} across both intrinsic (e.g., $\mathcal{M}_{\rm c}$, $q$, $a_{1,2}$) and extrinsic (e.g., $d_{\rm L}$, $\theta_{JN}$, $\psi$) parameters, demonstrating that the neural posterior estimator accurately reproduces the full Bayesian inference results at a fraction of the computational cost.
The true parameters (solid gray lines) are also well recovered within the inferred $68\%$ credible regions (shaded regions in dark blue; while $95\%$ credible regions are shared in light blue) with a sampling efficiency of $2.3\%$.
The Bayes factors obtained with \bilby{} and \dingolensing{} --- $\log_{10}\BFlens={19.0}$ and ${18.9}$, respectively --- are consistent within the statistical uncertainties, confirming the reliability of \dingolensing{} in evaluating the lensing hypothesis while reducing the inference time to $\sim40$~s.

\subsection{Recovery of GWs diffracted by point lenses}

\begin{figure}[t!]
    \centering
    \includegraphics[width=0.9\columnwidth]{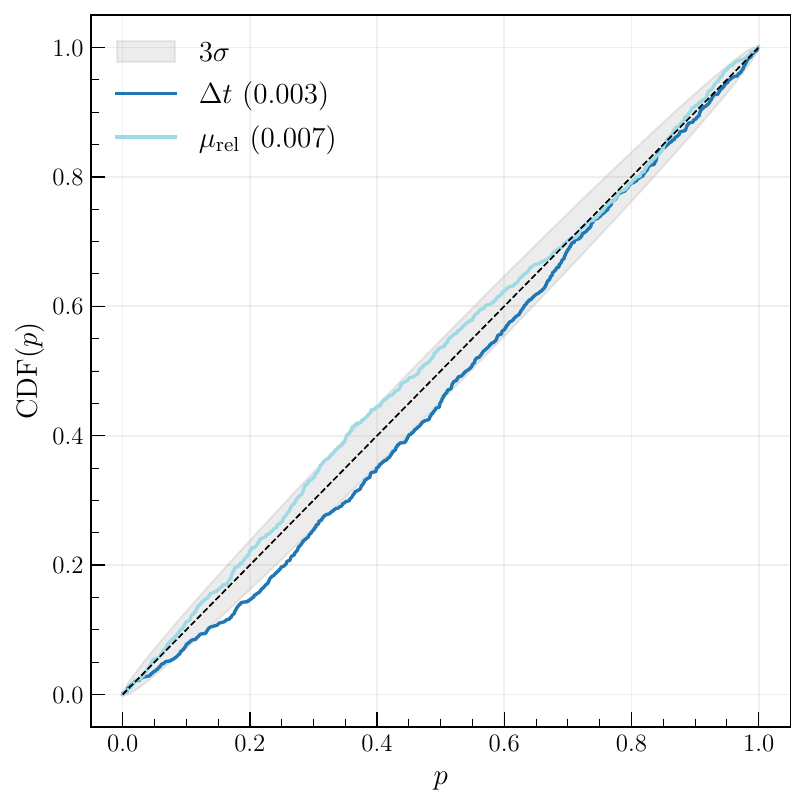}
~    \caption{P-P plot showing the calibration of the posterior distributions for the lensing parameters $\Dt$ and $\murel$, obtained from $10^3$ simulated injections of point-mass–lensed GW signals. Each curve compares the cumulative fraction of true parameter values enclosed within given credible levels against the ideal uniform expectation (black dashed line). The light-gray band indicates the expected $3\sigma$ statistical uncertainty for a perfectly calibrated inference. The analysis uses the geometrical optics approximation to recover signals generated with a point-lens model, demonstrating accurate recovery of both the time delay and relative magnification distributions.
}
    \label{pml_pp}
\end{figure}

To assess the effectiveness of the geometrical optics approximation in capturing lensing distortions, we perform inference on a lensed GW signal produced by a point-mass lens for which the amplification factor is analytical and depends on the redshifted lens mass $M_{\mathrm{Lz}}$ and impact parameter $y$~\cite{Takahashi:2003ix}. 
The point-mass lens always produces two images, but also includes diffraction effects which are not present in our parametrization of interfering chirps in (\ref{eq:F_two_images_point}).  
In practical terms, we use the implementation of the point-mass lens model coded in \modwaveforms, 
which is derived on Ref.~\cite{Ezquiaga:2020spg}. 
Within the geometric optics limit, the lens mass and impact parameter can be mapped one-to-one with the lensing time delay and relative magnification:
\begin{align}
    \Dt &= \frac{4GM_{\mathrm{Lz}}}{c^3}\Delta T(y)\,,\\
    \murel &= \left|\frac{y\sqrt{y^2+4}-(y^2+2)}{y\sqrt{y^2+4}+(y^2+2)}\right|\,,
\end{align}
where the dimensionless time delay is determined by the impact parameter only
\begin{equation}
    \Delta T(y) = \frac{1}{2}y\sqrt{y^2+4}-\ln\left|\frac{y-\sqrt{y^2+4}}{y-\sqrt{y^2+4}}\right|\,.
\end{equation}
The time delay scales linearly with the lens mass. 
For a 100$M_\odot$ lens, the time delay is on the order of $4GM_{\mathrm{Lz}}/c^3\approx 2$ms. 
In the limit of $y\to0$, then $\murel\to1$ and $\Dt\to0$.  

Figure~\ref{pml_pp} presents the P-P plot assessing the calibration of posterior inferences for the lensing parameters $\Dt$ and $\murel$, based on $10^3$ simulated injections of GW signals lensed by a point-mass lens, and $10^4$ sample were drawn from the lensed network for each injection.
We simulate lensed events  by uniformly sampling lens masses from $500 M_\odot$ to $1500M_\odot$. 
Impact parameters are drawn uniformly from $0$ to $1.5$. 
This choice of parameters ensures that the associated time delay does not exceed $0.1$s, which is the maximum time delay in the training set. 
Note that because the simulated signal's duration are always larger than 0.1s, we are testing examples in the wave-optics regime~\cite{Nakamura:1997sw,Takahashi:2003ix}.  
Each curve shows the cumulative fraction of true parameter values contained within the corresponding nominal credible intervals, compared to the ideal uniform expectation indicated by the black dashed line. 
The light gray region denotes the expected $3\sigma$ statistical variation for a perfectly calibrated inference under the null hypothesis. 
While both $\Dt$ and $\murel$ curves exhibit mild deviations beyond the $3\sigma$ envelope, these offsets are small and do not indicate significant systematic biases. Moreover, we also include the corresponding $p$-values from the Kolmogorov-Smirnov test in the legend.
This level of agreement demonstrates that, even when using the geometrical optics approximation to recover signals generated with a point-lens model, the inference remains well calibrated and the residual systematics are negligible relative to statistical fluctuations.

\begin{figure}[t!]
    \centering
    \includegraphics[width=0.9\columnwidth]{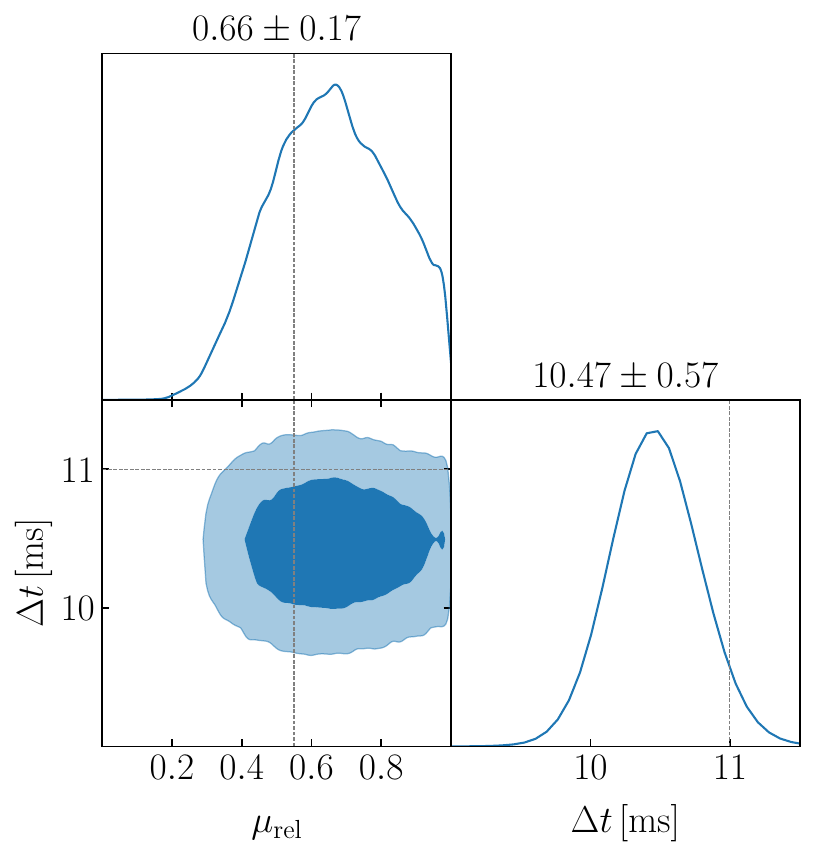}
~    \caption{The injection corresponds to a point-mass lens with $M_{\mathrm{Lz}} = 900M_{\odot}$ and $y = 0.3$. The inferred $\Dt = 10.47 \pm 0.57~\mathrm{ms}$ and $\murel = 0.66\pm0.17$ are consistent with the geometrical optics predictions of $\Dt = 11~\mathrm{ms}$ and $\murel = 0.5$, demonstrating the capability of the geometrical optics approximation to accurately reproduce lensing-induced distortions.}
    \label{corner_pml}
\end{figure}

To further illustrate the quality of individual recoveries, we focus on a representative example of a lensed signal with a duration of $\sim0.1\rm{s}$.
Figure \ref{corner_pml} presents a subset of the 17-dimensional corner plot and illustrates the recovery of a lensed GW signal using the \dingolensing{} framework for a point-mass lens with $M_{\mathrm{Lz}}=900~M_{\odot}$ and $y=0.3$.
$10^{5}$ samples have been drawn from the lensed network with utilizing importance sampling.
The two-dimensional posterior distributions show tight constraints on the lensing parameters, indicating that the model successfully captures the lens-induced distortions.
The inferred $\Dt=10.47 \pm 0.57~\mathrm{ms}$ and $\murel = 0.6 \pm 0.17$ are in excellent agreement with the geometrical optics predictions of $\Dt=11~\mathrm{ms}$ and $\murel = 0.5$, respectively, demonstrating the capability of the geometrical optics approximation to accurately reproduce the lensing effects encoded in the waveform even in the wave-optics regime.

We note that this demonstration does not rely on a fully self-consistent point-lensed source model for parameter estimation. Such models, while analytically tractable, neglect waveform modifications arising from finite-source effects and frequency-dependent interference, which become significant outside the geometrical optics regime.
Instead, our approach focuses on validating the ability of the \dingolensing{} network to recover the effective observables ($\Dt$, $\murel$) that characterize the lensing distortion within this limit.
A direct comparison with full Bayesian recovery using \bilby{} is not performed here, as the inference is expensive and other codes have already implemented this~\cite{Wright:2021cbn,gwmat}.
Our inference test therefore serves as a controlled and efficient validation of the capability of the network to reproduce the predictions of geometrical optics with high fidelity.

\subsection{Lensing detectability}

\dingolensing{} enables us to assess the detectability of gravitational waveform distortions induced by lensing. 
To that end, we simulate a large set of lensed and nonlensed events for our fiducial Hanford-Livingston O5 detector network. 
We sample events uniformly in a comoving volume between $1$Gpc and $15$Gpc for chirp masses between $30 M_\odot$ and $60 M_\odot$ and mass ratios larger than $0.3$. 
The rest of the parameters are sampled from the fiducial distributions used in network training. 
The reference trigger GPS time is fixed to $100$s (with a uniform scatter of $\pm0.1$s), and we only consider events with a network SNR larger than $8$. 
For every set of sampled parameters, we generate a frame file and define the detector strain data for a given Gaussian noise realization of the power spectral density. 
We analyze each frame file with both our lensed and nonlensed networks in order to compute the lensing Bayes factor using importance sampling.  
We have validated our simulations against \bilby{} for instances with different sampling efficiencies, see Appendix \ref{app:BF_sampling_efficiency}, finding good agreement in the Bayes factor for $\epsilon>10^{-4}$, which we set as a lower bound. 

%-----------------
%FIGURE
%-----------------
\begin{figure}[t!]
    \centering
    \includegraphics[width=\columnwidth]{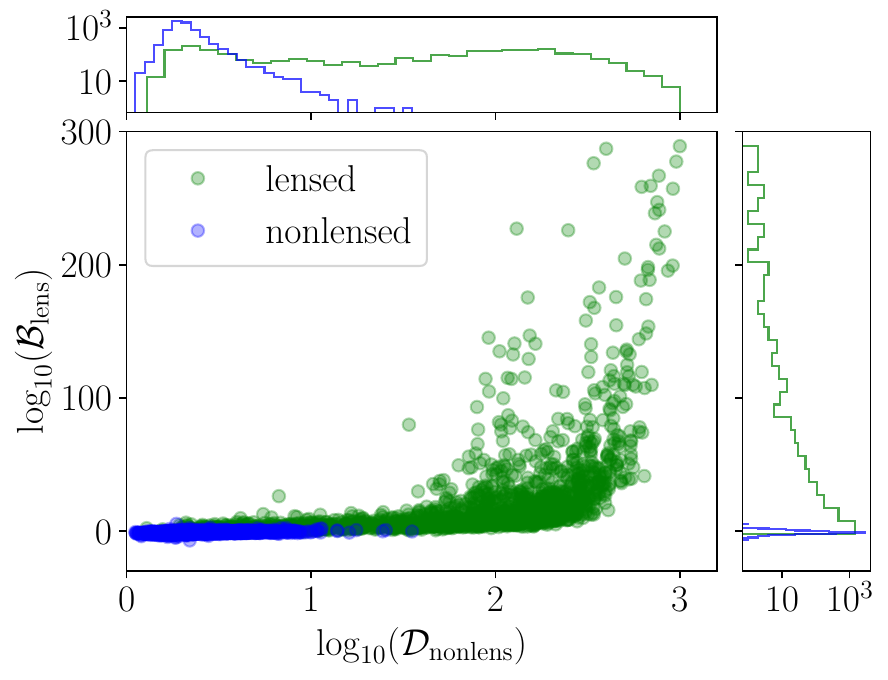}
    \caption{Distribution of lensing Bayes factors ($\BFlens$) and distance away from the nonlensing hypothesis ($\Dlens$) for the set of \Nforeground{} lensed and \Nbackground{} nonlensed simulated events considered in this work.}
    \label{fig:BLUvsDistance}
\end{figure}
%-----------------
%----------------- 

Lensing candidates can be identified through different methods. For example, using the posterior samples directly would give an indication of an inconsistency with the nonlensing hypothesis if the lensing parameters deviate away from $0$. 
To quantify this inconsistency, we follow a similar approach to Ref.~\cite{Ezquiaga:2023xfe} and compute the (Gaussian) distance to $(0,0)$ in the 2D parameter space $(\Dt,\murel)$. 
We define this distance away from the nonlensing hypothesis as $\Dlens$, see Appendix \ref{app:distance} for details. 
Such distance is weighted by the uncertainty of the lensing parameters. 
Well-constrained posteriors away from 0 would lead to larger distances. 
Because both lensing parameters are bounded at 0, distances are always larger than 1. 
We compare the distributions of $\Dlens$ and $\BFlens$ in Fig.~\ref{fig:BLUvsDistance} for \Nforeground{} lensed and \Nbackground{} nonlensed simulations. 
We find that both are correlated, with lensed simulations displaying a large tail at high values. 
Noticeably, the distribution of nonlensed simulations has a narrower range of values both in distance and Bayes factors.

\begin{figure}[t!]
    \centering
    \includegraphics[width=\columnwidth]{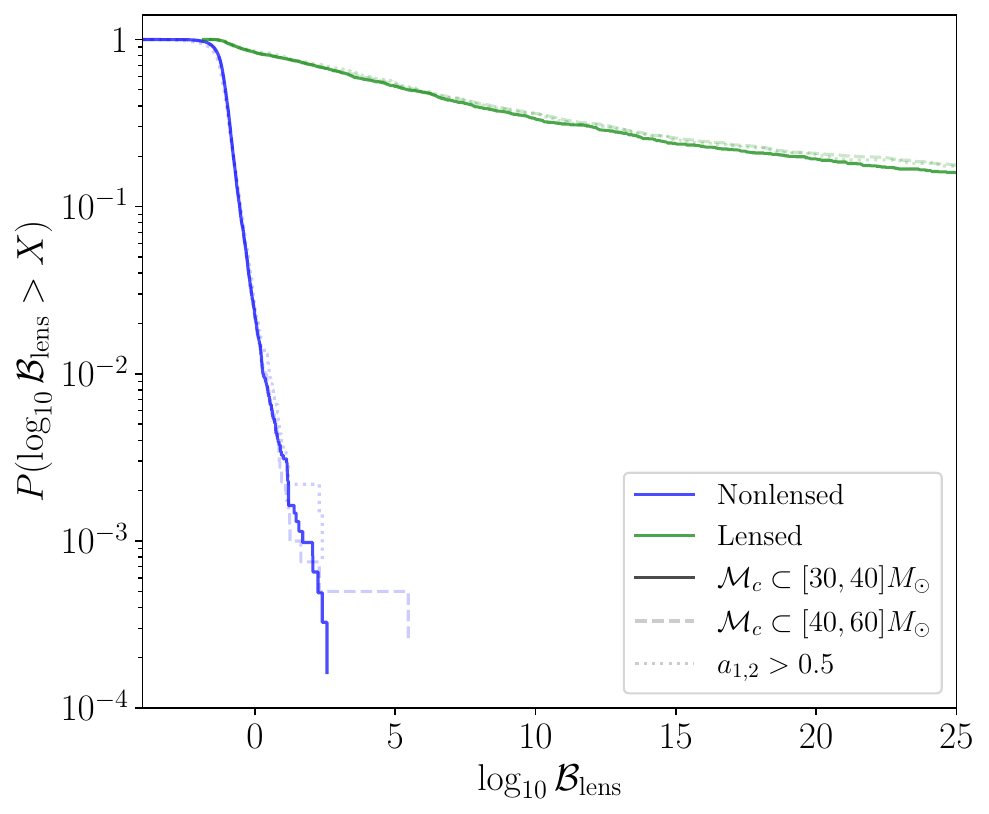}
    \caption{Complementary cumulative distribution function of lensing Bayes factors ($\mathcal{B}_\mathrm{lens}$) for simulated lensed and nonlensed with O5 sensitivity. 
    We distinguish between events with chirp masses between [30,40]$M_\odot$ (solid lines) and [40,60]$M_\odot$ (dashed lines). 
    For all chirp masses, we also plot the cases with both component dimensionless spin magnitudes smaller than 0.5 (dotted lines).}
    \label{fig:backgrounds}
\end{figure}

We explore in more detail the distribution of Bayes factors in Fig. \ref{fig:backgrounds}, where we plot the complementary cumulative distribution functions.
Lensed simulations display large Bayes factor, with \ForegroundLargerFive{} of them being larger than 5.  
However, for a given lensing candidate, its statistical significance depends on how it compares to a background, which defines the expectations of the nonlensed hypothesis. 
In other words, it is necessary to compute the \textit{false alarm probability}. 
Similar background studies with \dingo{} have also been performed in the literature when searching for eccentric signals~\cite{Gupte:2024jfe}.

We find that only \BackgroundLargerZero{} of nonlensed simulations favor lensing, $\BFlens>0$. 
The background, on the other hand, is sensitive to the underlying distributions being considered. 
We illustrate this in Fig.~\ref{fig:backgrounds} by distinguishing between nonlensed events with chirp masses in the ranges [30,40]$M_\odot$ and [40,60]$M_\odot$. 
We find that nonlensed events with larger masses tend to mimic lensing at a higher rate. 
If we consider only the subset of nonlensed simulations with either of the dimensionless, component spin magnitudes ($a_{1,2}$) larger than 0.5, we find that the false alarm rate also increases.  
This matches with the expectation that highly spinning signals could produce larger precession and be shorter in duration, being more likely to resemble lensing by random chance. 

%-----------------
%FIGURE
%-----------------
%-----------------
%----------------- 

%==========================================================
\section{Conclusions}\label{Section: Conclusion}
%==========================================================
In this work, we introduce \dingolensing, 
a simulation-based inference
framework for performing fast and accurate inference on gravitationally lensed GW signals with deep learning approaches. 
\dingolensing{} bypasses the computational bottlenecks of conventional samplers for Bayesian inference of distorted GW from lensing while maintaining statistical interpretability.

Using a dataset of $5\times10^6$ simulated waveforms, we have trained lensed and nonlensed models capable of producing well-calibrated posteriors, as validated through P-P plots and comparisons with standard parameter estimation analyses with \bilby~\cite{Ashton:2018jfp}. 
The results demonstrate excellent agreement across intrinsic, extrinsic, and lensing parameters, while reducing the inference time from weeks to tens of seconds per event. 

We demonstrate the potential of \dingolensing{} by training networks with LIGO detectors at design sensitivity. 
We find that the lensing parameters can be accurately recovered, with precision in the lensing time delay of the order of milliseconds. 
We show that our fiducial lens model with two overlapping chirps is able to recover lensed GWs diffracted by point-mass lenses. 
Moreover, we assess the detectability of lensed GWs by simulating thousands of lensed and nonlensed events. 
We find that for typical GW signals, the distribution of lensing Bayes factors of lenses and nonlensed events is clearly distinguishable.  
However, we also find
that the statistical significance of a given lensed event is sensitive to the background it is compared to. 
In particular, we observe a dependence in mass and spin. 
When comparing a signal with a background of heavier and more spinning binaries, the required %larger 
Bayes factors to arrive at the same false alarm probability is larger than when comparing with lighter and less spinning signals. % to arrive at the same false alarm probability. 
A detailed detectability forecast will be addressed in future work.  

The discovery of lensed GWs requires well-defined detection strategies. 
\dingolensing{} can contribute to them in two different stages.  
First, it can be used for the identification of lensed candidates with a set of networks trained on the average power spectral density over a given observing run. 
These networks should encompass broad mass ranges characterizing different signals duration. 
The maximum lensing time delay should be adjusted for each network so that there are single, distorted lensed waveforms. 
Lensing candidates can then be identified through two alternative methods. 
On one hand, one can compute the distance between the lensing posterior samples and the nonlensing hypothesis, $\Dt=\murel=0$ using only neural posterior estimation. 
On the other hand, through the use of importance sampling, the Bayes factor between the lensed and nonlensed hypothesis can be compared to pre-computed simulations of nonlensed events.
To avoid potential biases, such simulations should describe agnostic models, i.e. uniformly distributed priors in the networks, and astrophysically motivated ones, i.e. matching GW population models with lensing optical depths. 

Second, \dingolensing{} can be used 
to train specialized networks for the most significant candidates. 
This second step is essential to account for the nonstationarity of GW detector sensitivities and the best-fit waveform approximant and signal properties. 
The specialized network then provides the means to compute a dedicated background to assess the statistical significance of a particular lensed candidate, which would require thousands/millions of simulations to arrive at 3$\sigma$/5$\sigma$ detections. 
Obtaining a high statistical significance is a necessary step before drawing any astrophysical implication.

Altogether, \dingolensing{} 
provides a scalable pathway for rapid lensing analyses in upcoming LVK observing runs. 
Its efficiency enables large-scale background estimation and systematic population studies that are otherwise computationally prohibitive. 
Future work will extend this framework to more complex lens models, incorporate waveform systematics, and integrate \dingolensing{} into end-to-end pipelines for real-time identification and characterization of lensed GW events.
Non-gaussianity and nonstationarity in realistic noise can affect both the accuracy of NPE and the detectability of lensing.
While the impact on NPE accuracy remains unclear, lensing detectability is likely to be overestimated.
A systematic study of NPE performance and lensing detectability under different glitch mitigation techniques, such as \texttt{BayesWave}~\cite{Cornish:2014kda,Hourihane:2022doe} and \texttt{gwsubtract}~\cite{Davis:2018yrz,Davis:2022ird}, will be carried out in future work.

\begin{acknowledgements}
We thank Srashti Goyal, Miguel Zumalacárregui, Nihar Gupte, Annalena Kofler, Stephen R. Green, Alvin Li, Elwin Ka Yau Li, and Arthur Offermans for valuable discussions and insightful comments. 
We also thank David Keitel for carefully reviewing this article for LVK, providing many constructive comments. 
This work was supported by Research Grants No.~VIL37766 and No.~VIL53101 from Villum Fonden and the DNRF Chair Program Grant No.~DNRF162 by the Danish National Research Foundation.
This work has received funding from the European Union's Horizon 2020 research and innovation programme under the Marie Sk\l{}odowska-Curie Grant Agreement No.~101131233.  
J.~M.~E is also supported by the Marie Sk\l{}odowska-Curie Grant Agreement No.~847523 INTERACTIONS.
The Center of Gravity is a Center of Excellence funded by the Danish National Research Foundation under grant no.~DNRF184.
The Tycho supercomputer hosted at the SCIENCE HPC center at the University of Copenhagen was used for supporting this work. 

This research has made use of data or software obtained from the Gravitational Wave Open Science Center (gwosc.org), a service of the LIGO Scientific Collaboration, the Virgo Collaboration, and KAGRA. This material is based upon work supported by NSF's LIGO Laboratory which is a major facility fully funded by the National Science Foundation, as well as the Science and Technology Facilities Council (STFC) of the United Kingdom, the Max-Planck-Society (MPS), and the State of Niedersachsen/Germany for support of the construction of Advanced LIGO and construction and operation of the GEO600 detector. Additional support for Advanced LIGO was provided by the Australian Research Council. Virgo is funded, through the European Gravitational Observatory (EGO), by the French Centre National de Recherche Scientifique (CNRS), the Italian Istituto Nazionale di Fisica Nucleare (INFN) and the Dutch Nikhef, with contributions by institutions from Belgium, Germany, Greece, Hungary, Ireland, Japan, Monaco, Poland, Portugal, Spain. KAGRA is supported by Ministry of Education, Culture, Sports, Science and Technology (MEXT), Japan Society for the Promotion of Science (JSPS) in Japan; National Research Foundation (NRF) and Ministry of Science and ICT (MSIT) in Korea; Academia Sinica (AS) and National Science and Technology Council (NSTC) in Taiwan.
\end{acknowledgements}

\section*{Data availability}
\dingolensing{} is available on GitHub~\cite{dingo-lensing}. 
The neural networks used in this work will be made public upon publication. 

\textit{Note added.}---During the preparation of the manuscript, we learnt that Caldarola et al. have similarly explored using \dingo{} for analyzing signals lensed by a point mass~\cite{Caldarola:2025oxr}.

\appendix

\section{Nonlensed Network Validation}
\label{Appdx:not lensed}
We train the nonlensed network following the setup and prior distribution described in the main text and Table~\ref{tab:priors}, except that the lensing parameters are excluded from the prior. Our architecture comprises $\sim 10^8$ learnable parameters and $\sim 10^7$ fixed parameters.

\begin{figure}[h!]
    \centering
    \includegraphics[width=\columnwidth]{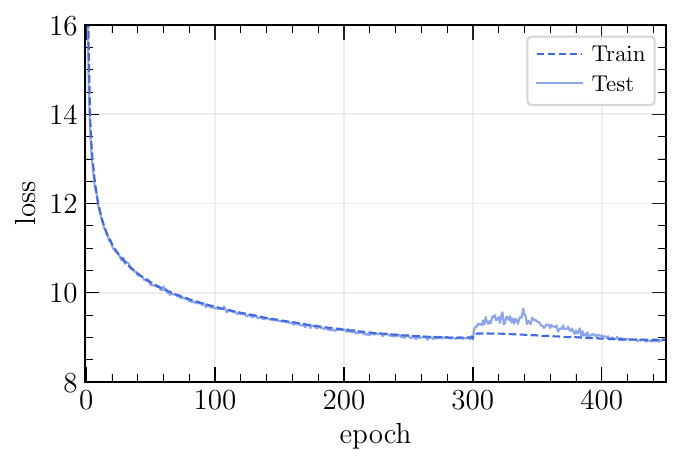}
    \caption{The loss as a function of the number of training epochs of the nonlensed network, shown for both the training set (dashed) and testing set (solid).}
    \label{loss_not_lensed}
\end{figure}

Figure~\ref{loss_not_lensed} presents the evolution of the loss function over the course of training.
Throughout optimization, the training (dashed) and validation (solid) losses remain closely aligned, indicating that the model generalizes well without signs of overfitting.
Both curves show a smooth, monotonic decline, reflecting a stable and well-behaved training process.

Around epoch~$300$, a distinct kink appears, corresponding to the transition to stage~$1$, when the previously frozen layers are unfrozen for fine-tuning.
The loss subsequently stabilizes after roughly $400$ epochs, implying convergence and the saturation of performance gains.
Training is therefore halted at epoch~$450$, once no appreciable improvement in the validation loss is observed.
The entire training procedure requires approximately $10$~days on an NVIDIA A100 PCIe GPU.
 
\begin{figure}[h!]
    \centering
    \includegraphics[width=0.9\columnwidth]{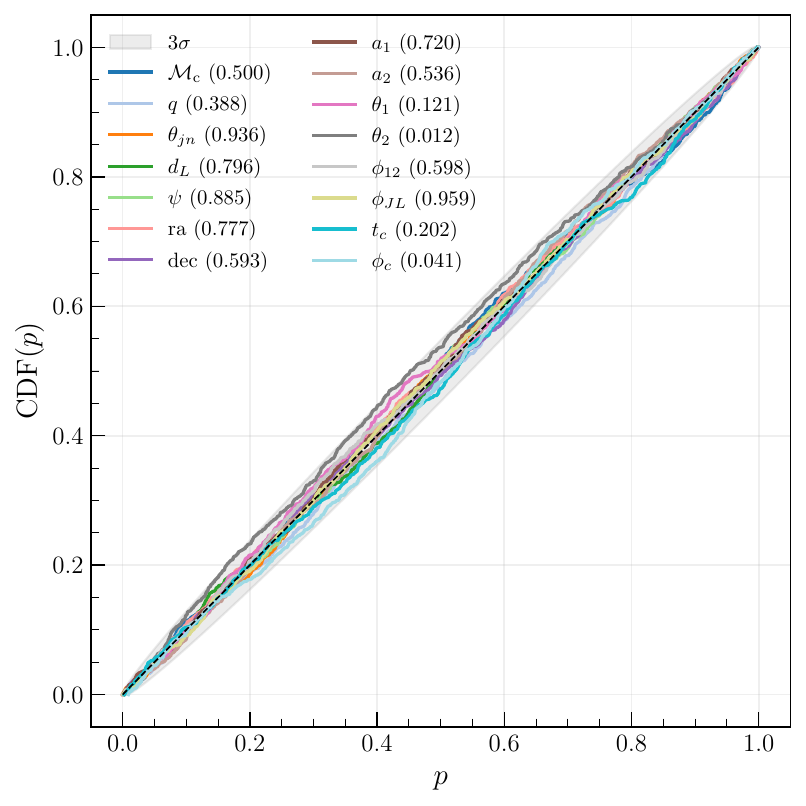}
    \caption{P-P plot for the nonlensed neural posterior estimation model constructed using $10^3$ unlensed injections and no importance sampling.
    For each injection, a posterior is generated, and the percentile value of the injected parameter is computed.
    Each colored line corresponds to the cumulative distribution function (CDF) of an individual parameter, with the Kolmogorov–Smirnov test $p$-values given in the legend.}
    \label{pp_plot_not_lensed}
\end{figure}

Figure~\ref{pp_plot_not_lensed} displays the P-P plot obtained from the posterior distributions of the nonlensed injection set.
The black diagonal indicates the ideal expectation that $x\%$ of the true parameter values fall within the corresponding $x\%$ credible intervals.
The colored curves representing individual parameters follow this diagonal closely, signifying that the inferred posteriors are statistically well calibrated.
Such consistency demonstrates that the inference network is accurately able to cover the true parameter space. 
Furthermore, as seen by the $p$-values reported in the legend, the Kolmogorov–Smirnov test confirms the reliability of the sampling procedure. 

\begin{figure*}[t!]
    \centering
    \includegraphics[width=\textwidth]{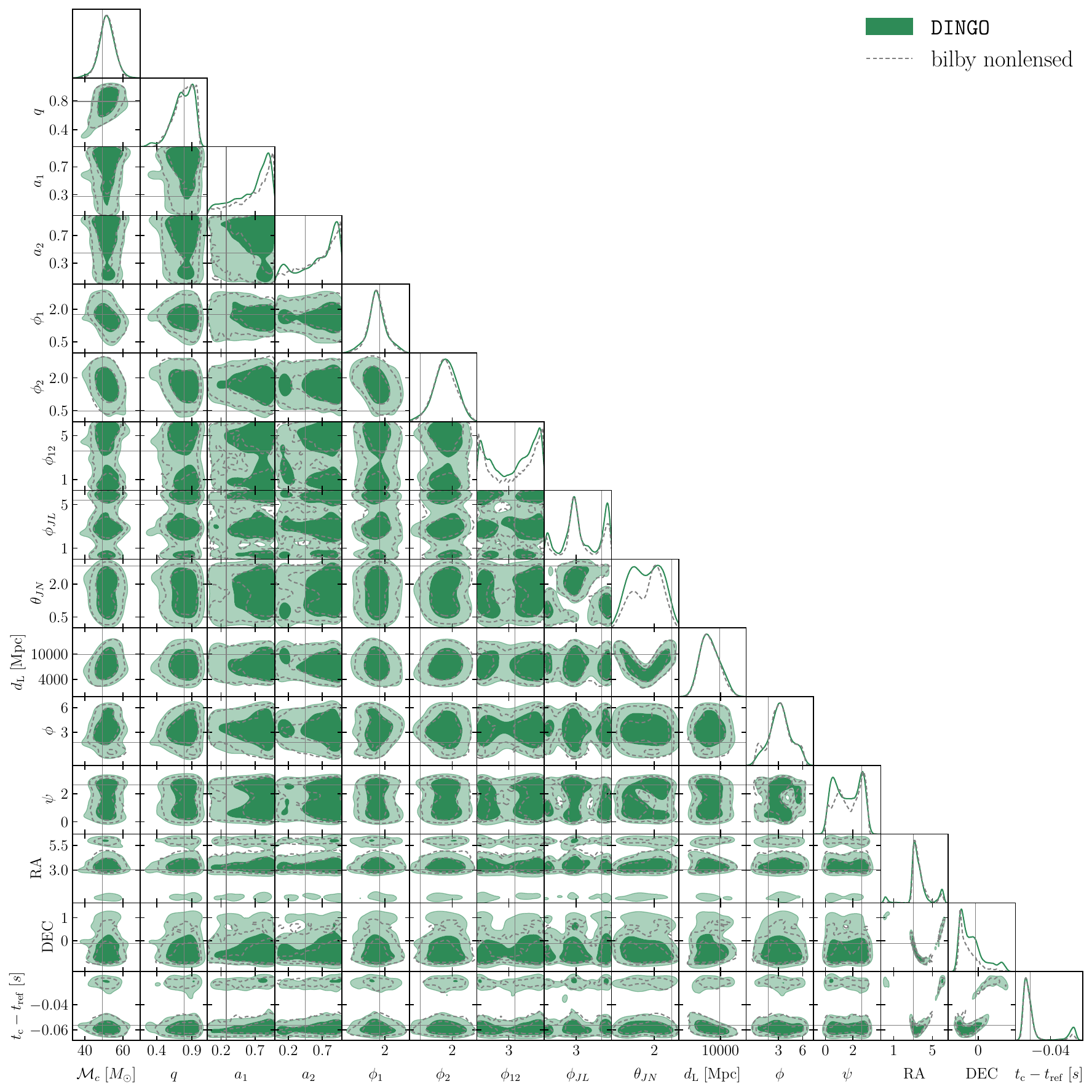}
    \caption{Posterior distribution of an injection recovered using \dingo{} (nonlensed network) (green solid curves) and \bilby{} (gray dashed curves), respectively. The two methods agree with each other and both recovered the injected values (solid gray lines; also tabulated in Table~\ref{tab:priors}) well.}
    \label{corner_nonlensed}
\end{figure*}

Figure~\ref{corner_nonlensed} shows a corner plot comparing the posterior distributions from the \dingo{} nonlensed model (solid green contours) with those from the reference \bilby{} analysis (dashed gray contours) for the same \emph{lensed} injection used in Sec.~\ref{subsec:comparison_with_bilby}.
The true parameters and the priors adopted in the \bilby{} run are listed in Table~\ref{tab:priors}.
The one- and two-dimensional marginalized posteriors exhibit excellent consistency between \dingo{} and \bilby{} across both intrinsic and extrinsic parameters, confirming that the neural posterior estimator faithfully reproduces the full Bayesian inference. 
The sampling efficiency for \dingo{} here is $0.3\%$. 

\begin{figure}
    %\centering
    \includegraphics[width=\columnwidth]{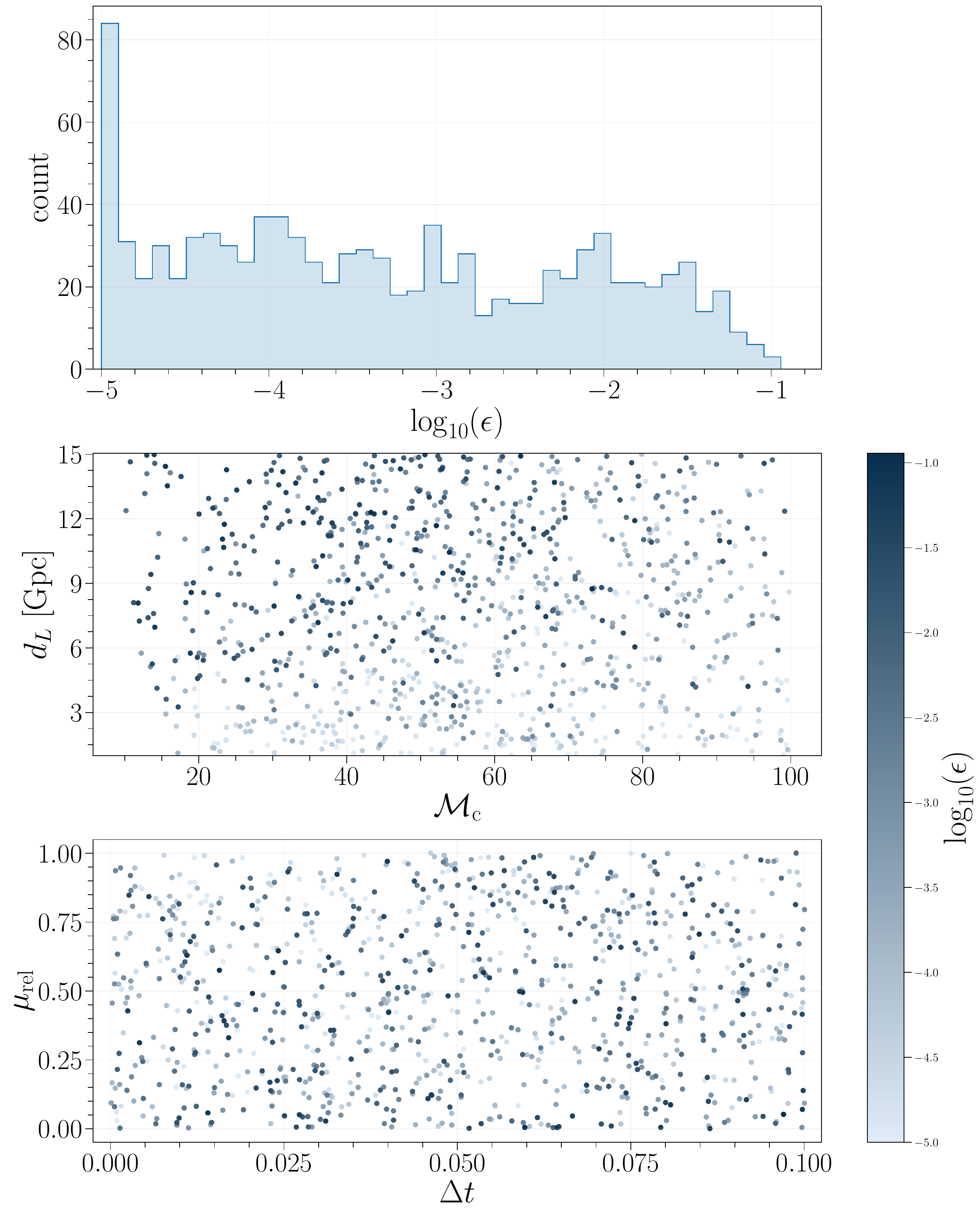}
    \caption{Distribution of the sampling efficiencies $\epsilon$ of $10^3$ lensed injections for the lensed network. $10^5$ samples are drawn for each injections The top panel shows the histogram of $\log_{10}(\epsilon)$, with most samples distributed between $10^{-4}$ and $10^{-1}$. The middle panel illustrates the dependence of $\epsilon$ on the source parameters, where each point represents an injection colored by $\log_{10}(\epsilon)$ in the $(\mathcal{M}_{\rm c}, d_L)$ plane. A weak trend is observed in which higher $\mathcal{M}_{\rm c}$ and larger $d_L$ correspond to an improved sampling efficiency. The bottom panel shows the variation of $\epsilon$ with the lensing parameters $(\Dt, \mu_{\mathrm{rel}})$, indicating no strong correlation but a broad spread consistent with the added complexity of the lensing parameter space.
    }
    \label{sample_effiency}
\end{figure}

\begin{figure}
    %\centering
    \includegraphics[width=\columnwidth]{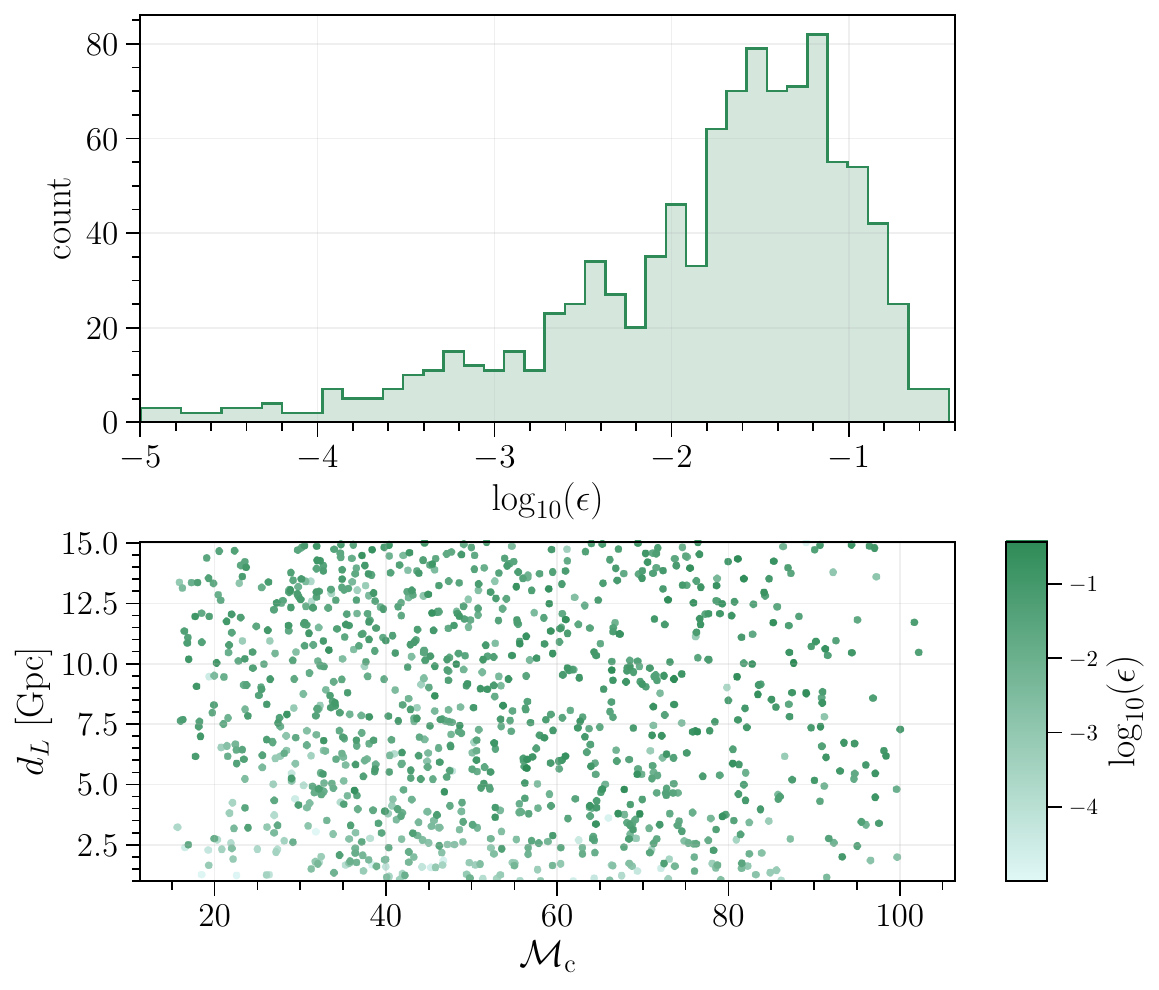}
    \caption{Distribution of the sampling efficiency $\epsilon$ of $10^3$ nonlensed injections for the nonlensed network. $10^5$ samples are drawn from the nonlensed network. The top panel shows the histogram of $\log_{10}(\epsilon)$, illustrating that most posterior samples have efficiencies between $10^{-2.5}$ and $10^{-1}$.  The bottom panel presents the relation between luminosity distance $d_L$ and chirp mass $\mathcal{M}_{\rm c}$, color-coded by $\log_{10}(\epsilon)$.  The results indicate that lower-efficiency samples are more prevalent at small $d_L$ and moderate $\mathcal{M}_{\rm c}$, while the overall spread remains broad across the parameter space.
    }
    \label{sample_effiency_notlensed}
\end{figure}

\section{Bayes factor dependence on the sampling efficiency}
\label{app:BF_sampling_efficiency}

To further evaluate the efficiency of the posterior reconstruction, we examine the sampling efficiency $\epsilon$ obtained after applying importance sampling in our simulations validating the lensed and nonlensed networks.
We perform inference on $10^3$ lensed injections and $10^5$ samples was drawn for each injection from the lensed network.
The results for the lensed network are 
shown in Fig.~\ref{sample_effiency}.
The top panel shows the histogram of $\log_{10}(\epsilon)$, with samples distributed between $10^{-1}$ and $10^{-4}$. 
The middle panel displays the distribution in the $(\mathcal{M}_{\rm c}, d_{\rm L})$ plane, which follows a trend of higher $\epsilon$ at larger $d_{\rm L}$, indicating that the lensed network learns more effectively at lower SNR. 
This is expected as posteriors become broader and it is easier for distributions to overlap.
The bottom panel demonstrates the distribution of $\log_{10}(\epsilon)$ in the lens parameter space and reveals that the $\log_{10}(\epsilon)$ shows no significant dependence on the lensing parameters. 

Figure~\ref{sample_effiency_notlensed} presents the distribution of the $\epsilon$ obtained from the nonlensed inference network.
We perform inference $10^3$ nonlensed injections and $10^5$ samples was drawn for each injection from the nonlesed network.
The upper panel shows the histogram of $\log_{10}(\epsilon)$, indicating that most posterior samples achieve efficiencies between $10^{-2.5}$ and $10^{-1}$, with a tail extending toward lower values. 
The lower panel illustrates how the efficiency varies across the source parameter space, where each point corresponds to an injection colored by $\log_{10}(\epsilon)$ in the $(\mathcal{M}_{\rm{c}}, d_L)$ plane. 
A mild trend is observed in which higher sampling efficiency is associated with systems of larger chirp mass and greater luminosity distance, suggesting that the nonlensed network more effectively reconstructs at lower-SNR and massive binaries. 
Overall, the nonlensed network exhibits stable performance across the population, with variations in $\epsilon$ primarily reflecting differences in signal strength rather than systematic inference bias.

Our main metric to evaluate lensed candidates is the lensing Bayes factor $\BFlens$. 
To validate the performance of our network, we randomly select  lensed and nonlensed events from our sets of simulations with different \dingolensing{} Bayes factors and sampling efficiencies and compare them with \bilby{}. 
We focus on cases with $\BFlens>0$. 
Simulated events with \dingolensing{} have $5\times10^4$ samples. 
The results are summarized in Table~\ref{tab:nonlensed_sims_BLU}. 
We observe that the Bayes factors are consistent with each other within statistical uncertainties for $\epsilon > 10^{-4}$. 
Therefore, we cut our simulations below that value, to end up with \Nforeground{} and \Nbackground{} lensed and nonlensed simulations, respectively.  

\begin{table}[htbp]
\centering
\begin{tabular}{lcccc}
\hline
ID & $\log_{10}\mathcal{B}_\mathrm{lens}^{\dingo}$ & $\log_{10}\mathcal{B}_\mathrm{lens}^{\bilby}$ & $\epsilon_\mathrm{lens}\times10^{3}$ & $\epsilon_\mathrm{nonlens}\times10^{3}$ \\
\hline
6 & 1.24 & -0.13 & 0.02 & 0.62 \\
594 & 0.07 & 0.10 & 0.12 & 8.43 \\
779 & 1.20 & 1.83 & 25.29 & 106.16 \\
1141 & 1.24 & 1.73 & 60.78 & 95.77 \\
1154 & 0.16 & 0.81 & 2.99 & 0.96 \\
1762 & 1.71 & 2.21 & 0.82 & 7.80 \\
3092 & 0.79 & 1.39 & 1.11 & 2.36 \\
\hline
19 & 144.25 & 117.34 & 0.03 & 0.02 \\
78 & 4.80 & 4.97 & 0.03 & 0.08 \\
241 & 277.40 & 277.45 & 0.84 & 0.20 \\
630 & 4.56 & 4.30 & 0.02 & 0.28 \\
654 & 2.69 & 2.85 & 0.34 & 19.83 \\
808 & 8.43 & 8.67 & 0.83 & 1.34 \\
817 & 11.90 & 12.37 & 1.04 & 8.51 \\
823 & 10.30 & 8.25 & 1.10 & 0.03 \\
873 & 10.43 & 11.35 & 0.07 & 0.03 \\
937 & 1.18 & 1.45 & 10.85 & 9.96 \\
965 & 4.51 & 4.91 & 23.49 & 31.39 \\
\hline
\end{tabular}
\caption{Comparison of the lensing Bayes factor $\BFlens$ computed with \dingolensing{} and \bilby{} for nonlensed and lensed simulated event with different sampling efficiencies $\epsilon$. 
We report both the sampling efficiency obtained with the lensed ($\epsilon_\mathrm{lens}$) and nonlensed ($\epsilon_\mathrm{nonlens}$) networks.}
\label{tab:nonlensed_sims_BLU}
\end{table}

\begin{table*}[htbp]
\centering
\begin{tabular}{llllllllllllllllll}
\hline
ID & $\mathcal{M}_\mathrm{c}/M_\odot$ & $q$ & $a_1$ & $a_2$ & $\phi_1$ & $\phi_2$ & $\phi_{12}$ & $\phi_{JL}$ & $d_\mathrm{L}$ [Mpc] & $\theta_{JN}$ & $\psi$ & $\phi_\mathrm{ref}$ & t$_\mathrm{ref}$ [s] & RA & DEC & $\Delta t$ [s] & $\mu_\mathrm{rel}$ \\
\hline
6 & 31.32 & 0.43 & 0.05 & 0.15 & 2.18 & 2.31 & 3.26 & 2.15 & 2098.68 & 0.36 & 2.26 & 0.69 & 99.92 & 2.73 & -0.60 & 0.00 & 0.00 \\
594 & 34.16 & 0.42 & 0.43 & 0.34 & 0.83 & 2.29 & 3.64 & 4.25 & 6192.94 & 0.59 & 1.09 & 4.67 & 100.10 & 3.79 & 0.99 & 0.00 & 0.00 \\
779 & 35.06 & 0.57 & 0.10 & 0.44 & 1.58 & 1.73 & 0.01 & 4.60 & 4038.44 & 1.63 & 1.60 & 3.22 & 99.99 & 3.87 & -1.10 & 0.00 & 0.00 \\
1141 & 52.37 & 0.39 & 0.04 & 0.11 & 1.69 & 0.33 & 4.46 & 5.86 & 3538.84 & 1.01 & 0.10 & 4.64 & 99.96 & 1.73 & 0.84 & 0.00 & 0.00 \\
1154 & 33.58 & 0.96 & 0.48 & 0.10 & 1.12 & 1.13 & 5.87 & 2.30 & 10810.94 & 2.78 & 2.38 & 2.82 & 99.99 & 0.48 & 1.20 & 0.00 & 0.00 \\
1762 & 32.98 & 0.72 & 0.46 & 0.36 & 1.13 & 0.71 & 0.60 & 3.74 & 3771.03 & 1.91 & 2.90 & 4.56 & 100.04 & 2.16 & 1.40 & 0.00 & 0.00 \\
3092 & 37.02 & 0.42 & 0.01 & 0.00 & 0.89 & 1.69 & 0.46 & 2.25 & 2125.64 & 2.35 & 0.06 & 0.12 & 99.99 & 1.49 & -0.27 & 0.00 & 0.00 \\
\hline
19 & 37.95 & 0.40 & 0.35 & 0.08 & 0.23 & 2.00 & 4.52 & 4.52 & 3608.73 & 2.41 & 1.32 & 0.40 & 100.02 & 3.19 & -0.84 & 0.07 & 0.94 \\
78 & 37.14 & 0.79 & 0.23 & 0.35 & 1.99 & 0.35 & 3.75 & 0.27 & 14037.69 & 2.68 & 0.39 & 1.55 & 99.90 & 0.07 & 1.04 & 0.04 & 0.44 \\
241 & 51.96 & 0.73 & 0.34 & 0.17 & 0.84 & 1.25 & 0.14 & 4.73 & 2047.32 & 0.62 & 2.80 & 0.26 & 99.99 & 5.84 & 0.69 & 0.10 & 0.30 \\
630 & 33.38 & 0.83 & 0.42 & 0.36 & 1.79 & 2.20 & 4.92 & 6.07 & 8883.59 & 2.87 & 2.04 & 5.37 & 100.04 & 4.31 & 1.05 & 0.03 & 0.84 \\
654 & 43.40 & 0.86 & 0.16 & 0.04 & 0.72 & 1.53 & 5.24 & 4.67 & 6622.99 & 0.97 & 1.25 & 1.90 & 99.91 & 3.08 & 1.49 & 0.09 & 0.52 \\
808 & 36.66 & 0.52 & 0.17 & 0.47 & 0.58 & 1.48 & 2.26 & 1.31 & 9722.49 & 2.78 & 0.79 & 6.03 & 100.03 & 0.01 & 0.47 & 0.02 & 0.81 \\
817 & 32.48 & 0.68 & 0.44 & 0.39 & 1.33 & 2.10 & 3.87 & 3.56 & 3721.55 & 1.36 & 1.88 & 2.22 & 100.09 & 1.47 & -1.13 & 0.04 & 0.52 \\
823 & 34.49 & 0.64 & 0.45 & 0.06 & 0.24 & 1.53 & 2.45 & 5.01 & 6490.86 & 2.82 & 1.61 & 3.77 & 99.96 & 4.20 & 0.77 & 0.01 & 0.94 \\
873 & 30.34 & 0.52 & 0.43 & 0.13 & 1.25 & 0.84 & 5.48 & 2.06 & 1710.62 & 1.71 & 2.36 & 3.37 & 100.07 & 3.22 & -0.25 & 0.02 & 0.73 \\
937 & 43.38 & 0.63 & 0.37 & 0.50 & 0.77 & 1.79 & 4.01 & 1.48 & 6277.43 & 1.39 & 2.66 & 0.28 & 100.06 & 5.93 & 0.29 & 0.03 & 0.30 \\
965 & 53.73 & 0.46 & 0.24 & 0.28 & 1.02 & 1.30 & 2.29 & 2.62 & 8872.92 & 0.65 & 2.77 & 4.32 & 99.94 & 3.41 & -0.55 & 0.08 & 0.48 \\
\hline
\end{tabular}
\caption{Parameters for the simulated events in Table \ref{tab:nonlensed_sims_BLU}.}
\label{tab:nonlensed_sims_params}
\end{table*}

During this validation, we have observed that the networks sometimes struggle to compute the evidences when the simulated mass ratios are close to the lower bound of the training set, $q=0.125$. 
This may be due to a difficulty in learning the sharp features of a railing posterior distribution.  
For this reason, we conservatively limit our background/foreground simulations to $q>0.3$ where we see good agreement as demonstrated above. 
The specific parameters of our validation set are presented in Table \ref{tab:nonlensed_sims_params}. 

\section{Distance away from the nonlensing hypothesis}
\label{app:distance}

For our lens model, cf. Eq. (\ref{eq:F_two_images_point}), a nonlensed event is characterized by having $\Dt=\murel=0$. 
Therefore, one can quantify the inconsistency with this hypothesis by measuring the distance of the lensing posterior samples away from this point. 
We define this quantity as $\Dlens$. 
In order for this distance to be meaningful, it needs to be normalized by the uncertainty of the parameters themselves. 
Approximating the posterior distributions as Gaussian, the distance in units of $\sigma$ is given by~\cite{Ezquiaga:2023xfe}
\begin{equation}\label{eq:gaussian_distance}
    D(\theta_1,\theta_2) = \sqrt{(\theta_1-\theta_2))^T (C_1+C_2)^{-1} (\theta_1-\theta_2)}\,,
\end{equation}
where $\theta_{1}$ and $\theta_{1}$ are the parameters for which the distance is computed, and $C_1$ and $C_2$ are their covariances. 
In this case, $\theta_1 =\{\Dt,\murel\}$ and $\theta_2=\{0,0\}$, defining 
\begin{equation}
    \Dlens=D(\{\Dt,\murel\},0)\,.
\end{equation} 
Since the prior in $\{\Dt,\murel\}$ is bounded at 0, the lensing distance satisfies $\Dlens\gtrsim 1$. 
It is to be noted that for non-Gaussian posteriors, the Gaussian approximation will typically lead to smaller distances. 
Therefore, $\Dlens$ is a conservative metric to find lensed candidates. 
The advantage of $\Dlens$ over the lensing Bayes factor $\BFlens$ is that it does not require importance sampling and can be obtained directly from neural posterior estimation with \dingolensing{}. 

\normalem
\bibliographystyle{apsrev4-1}
\bibliography{references}

\end{document}